\documentclass[12pt]{iopart}
\usepackage{graphics, iopams, mhequ}

\newtheorem{proposition}{Proposition}[section]
\newtheorem{definition}{Definition}[section]
\newtheorem{corollary}{Corollary}[section]
\newtheorem{theorem}{Theorem}
\newtheorem{lemma}{Lemma}[section]

\begin{document}
\title{Sub-exponential mixing of random billiards driven by thermostats}
\author{Tatiana Yarmola}
\address{Department of Theoretical Physics, \\
University of Geneva, \\
Ecole de Physique, DPT \\
24 quai Ernest-Ansermet, \\
1211 Gen\`{e}ve 4, Switzerland \\
\ead{tatiana.yarmola@unige.ch}}

\begin{abstract}
We study the class of open continuous-time mechanical particle systems introduced in the paper by Khanin and Yarmola \cite{Khanin}. Using the discrete-time results from \cite{Khanin} we demonstrate rigorously that, in continuous time, a unique steady state exists and is sub-exponentially mixing. Moreover, all initial distributions converge to the steady state and, for a  large class of initial distributions, convergence to the steady state is sub-exponential. The main obstacle to exponential convergence is the existence of slow particles in the system.
\end{abstract}


\section{Introduction}
Non-equilibrium steady states for open mechanical particle systems have been extensively studied in recent years as a part of the efforts to derive the Fourier Law from the laws of microscopic dynamics \cite{Balint,Collet,Jacquet,Nonequilibrium,Klages_Nicolis_Rateitschak,Khanin,Larralde_Leyvraz_MejiaMonasterio,Lin,Yarmola}. The existence of non-equilibrium steady states has been rigorously shown for only a handful of systems, mostly by using either spectral-theoretic arguments \cite{Anharmonic_EH,Anharmonic_EPR,Grigo} or Harris' ergodic theorem \cite{Khanin, AnharmonicExponential}; these methods automatically yield exponential mixing and exponential rates of convergence of initial distributions to the unique steady state. For systems that do not mix exponentially fast, which is frequently the case for continuous-time mechanical particle systems driven by heat reservoirs, general methods to prove the existence of non-equilibrium steady states are virtually nonexistent. In some special cases, one may show tightness using direct estimates on the return times to the heat reservoirs \cite{GLP, Yarmola2}, and, provided that the heat reservoirs randomize velocities fully, show the existence of the steady states. A great exposition of ideas and difficulties associated with this task is presented in \cite{GLP}. This paper, together with \cite{Khanin}, provides a rigorous proof of the existence of a unique steady state with sub-exponential mixing rates for a class of random billiards driven by thermostats which randomize the velocities only in one direction, leaving a large memory trace in the system. In addition, for a large class of initial distributions, we obtain polynomial lower bounds on the rates of convergence of to the steady state.

In the system we consider, the particles move freely between collisions with non-intersecting disk-shaped random scatterers arranged on a torus so that no particle can move indefinitely without colliding with a scatterer. Upon a collision of a particle with a scatterer, an energy exchange occurs, in which the tangential component of the particle's velocity is randomized according to a Gaussian distribution with variance proportional to the temperature assigned to the random scatterer, and the normal component of the velocity changes sign.

The disk-shaped random scatterers act as thermostats in the following sense:
Assume we set all the random scatterers at temperature $T$ and start with an ensemble of $N$ particles at temperature $T_0 \ne T$, i.e., the probability that a given particle has kinetic energy within $dE$ near $E$ is approximately $c e^{-\beta_0 E}dE$, which corresponds to the Gibbs distribution at temperature $T_0=\frac{1}{\beta_0}$. We show, inter alia, that the distribution of the particle kinetic energies converges to the Gibbs distribution at temperature $T$. Heat conduction in this system may be studied by assigning one thermostat a temperature $T_1$ different from $T$ assigned to all others. Further understanding of the properties of the non-equilibrium steady states is required for this task; we leave heat conduction studies of this system for future work.

The particles in this system do not interact, so the analysis can be simplified by considering a system with only one particle. In this situation, it is convenient to study the discrete dynamics of particle collisions with the boundaries of the thermostats. Upon collisions with thermostats, the normal component of the particle's velocity represents the memory kept in the system, while the tangential component gets fully randomized. Therefore, one can define a discrete Markov chain in position and normal velocity variables that represents the original dynamics at collisions. For this Markov chain, the existence and uniqueness of the invariant probability measure, which describes the steady state, along with exponential rates of convergence of the initial distributions to the invariant measure were obtained in \cite{Khanin} via Harris' ergodic theorem.

In this paper we extend the discrete-time results of \cite{Khanin} to continuous time. In section \ref{sect:settings and results} we introduce the model and state the results. We present the proof of the main theorem in section \ref{sect: main}. Harris' ergodic theorem does not directly apply to the continuous time system, since the system does not mix exponentially fast. However, one of the key observations in our proof is that the potential (Lyapunov) function and the minorization condition required for application of Harris' ergodic theorem in the discrete case can be used to obtain upper and lower bounds on certain integrals and the measures of certain sets.

Before proceeding with the actual proof we briefly review the results in \cite{Khanin} in \ref{subsect: review}. In \ref{subsect: finiteness} we prove the existence of a unique invariant probability measure (the steady state) by constructing a suspension flow over the discrete dynamics and showing that the average of the roof function is finite.  The mixing and convergence of initial distributions to the invariant measure are obtained by applying \cite[Theorem 6.1]{MeynII} by Meyn and Tweedie for general state Markov chains. The conditions of the theorem are ensured by adapting some of the arguments in \cite{Khanin} to continuous time setting. We outline the modifications required and we refer the reader to \cite{Khanin} for more details. In this sense this paper serves as a companion paper to \cite{Khanin}.

The main reason for sub-exponential rates of mixing is that, from time to time, the particle acquires low speed, which leads to long traveling times between collisions with thermostats. In \ref{subsect: not exponential} we show that the (invariant) measure of the set of particles that would not collide with a thermostat in time $\tau$ is approximately of the order of $1/\tau^2$. Similar rates have been observed for the system in \cite{Lefevere}. We present a proof that such abundance of the slow particles in the system implies that mixing rates are, at best, polynomial.

\section{Settings and Results} \label{sect:settings and results}

\begin{figure} \label{fig: configuration}
  \centering
  \resizebox{0.4\textwidth}{!}{
  \includegraphics{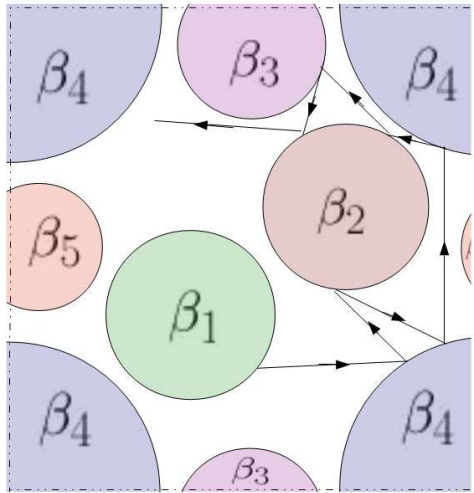}}
  \caption{Geometric Configuration}\label{fig: configuration}
\end{figure}

Let $\Gamma = \mathbb{T}^2 \setminus \cup_{i=1}^p D_i$ be a bounded horizon billiard table with a finite number of non-intersecting circular scatterers $D_1, \cdots, D_p$ with radii $R_1, \cdots, R_p$. Fig.\ref{fig: configuration} illustrates an example of such a configuration. Non-interacting particles move freely in $\Gamma$ with constant velocities between collisions with the scatterers. Each scatterer $D_i$, $1 \leq i \leq p$, is assigned a certain parameter $\beta_i$ which represents the inverse of its temperature. The scatterers play the role of thermostats.

Upon the collision of a particle with a thermostat $D_i$, let $v^-=(v_\perp^-, v_t^-)$ be the decomposition of the particle's velocity into its normal and tangential components with respect to the boundary $\partial D_i$. After the collision, the normal component of the velocity changes sign, so that $v_\perp^+=-v_\perp^-$, while the tangential component is absorbed by the thermostat and a new tangential component $v_t^+$ is drawn from the distribution $\sqrt{\frac{\beta_i}{\pi}} e^{-\beta_i v_t^2} dv_t^+$. The outgoing velocity has the decomposition $v^+=(v_\perp^+,v_t^+)$.

Since particles do not interact with each other, we can reduce our discussion to studying a system of only one particle. The phase space of such a system is $$\Omega=\{\mathbf{z}=(x,v):x \in \Gamma, v \in \mathbb{R}^2 \setminus \{0\}\}/\sim$$
where $\sim$ is the identification of points on the collision manifold that correspond to incidence and reflection with the choice of the incoming velocities $v^-=(v_\perp^-,v_t^-)$. Note that if the particle originates with $v \ne 0$, the probability to acquire zero velocity is $0$.

The dynamics of the system is  thus described by a Markov Process $\Phi_t$ on $\Omega$, which is deterministic between collisions with thermostats with random kicks at the moments of collisions with $\partial \Gamma$. Let $\mathcal{P}^t$ denote the transition probability kernel for $\Phi_t$.

\begin{definition}
A measure $\nu$ on $\Omega$ is called invariant under $\Phi_t$ if for any Borel set $A \subset \Omega$ and $t \geq 0$
$$\nu(A)=(\mathcal{P}_*^t \nu )(A) := \int_\Omega \mathcal{P}^t(\mathbf{z},A) d \nu (\mathbf{z}).$$
\end{definition}

\begin{definition}
The Markov process $\Phi_t$ is called mixing if an invariant probability $\nu$ exists and if for any Borel set $A \subset \Omega$,
$$ \sup\limits_{B \in \mathcal{B}} |\int_A \mathcal{P}^t(\mathbf{z},B)d\nu -\nu(A)\nu(B)| \to 0 \;\; \mathrm{ as } \;\; t \to \infty,$$
where $\mathcal{B}$ denotes the Borel sigma algebra on $\Omega$. \\
We say that $\nu$ is mixing exponentially fast if there exist $C,\alpha >0$ such that
$$\sup\limits_{B \in \mathcal{B}}|\int_A \mathcal{P}^t(\mathbf{z},B)d\nu -\nu(A)\nu(B)| \leq Ce^{-\alpha t}.$$
\end{definition}

\begin{definition}
An initial distribution $\lambda$ is said to converge to the invariant probability measure $\nu$ if
$$\lim\limits_{t \to \infty}\|\mathcal{P}^t\lambda - \nu\|=\lim\limits_{t \to \infty} \sup_{B \in \mathcal{B}}|\mathcal{P}^t\lambda(B) -\nu(B)| =0,$$
and does so exponentially fast if there exist $C,\alpha >0$ such that $\|\mathcal{P}^t\lambda - \nu\| \leq Ce^{-\alpha t}$. The norm $\|\cdot\|$ is called the total variation norm.
\end{definition}

The main result of this paper is as follows.

\begin{theorem} \label{main}
There exists a unique invariant probability measure for the continuous Markov process $\Phi_t$, which is absolutely continuous with respect to Lebesgue measure. This measure is mixing, but not exponentially mixing. All initial probability distributions converge to the invariant measure, while convergence rates are at most polynomial for a large class of initial distributions.
\end{theorem}

\section{Proof of Theorem \ref{main}} \label{sect: main}

As a preparation for the proof, we outline the settings and the results of \cite{Khanin} in \ref{subsect: review}. In addition, we mention several important ideas from the proof of the main theorem in \cite{Khanin} and state two propositions, to be used later, whose proofs are natural modifications of the proofs in \cite{Khanin}. In \ref{subsect: finiteness} we prove existence of an absolutely continuous (w.r.t. Lebesgue) invariant probability measure for the Markov process $\Phi_t$. The uniqueness is guaranteed by absolute continuity and one of the aforementioned propositions. In \ref{subsect: ergodicity} we show that the invariant measure is mixing and all the initial distributions converge to it. In \ref{subsect: not exponential} we demonstrate a class of initial distributions for which the rates of convergence to the invariant measure are sub-exponential.

\subsection{Review of settings and results of \cite{Khanin}} \label{subsect: review}
To study the continuous-time Markov process $\Phi_t$ it is convenient to consider a discrete in time dynamical system of particle collisions with the boundary $\partial \Gamma$ of $\Gamma$. When specifying $\Phi_t$ at collision points, we chose to keep old, incoming velocities. This choice enables a $2$-dimensional description of the discrete dynamics characterized by the collision points $r_n \in \partial \Gamma$ (parameterized by arc-length) and the absolute values of the normal velocity $v_\perp(n)$ at the moment of collision. Indeed, using a randomly generated tangential velocity $v_t(n)$ drawn from the probability distribution $\sqrt{\frac{\beta_i}{\pi}}e^{-\beta_i v_t(n)^2}d v_t(n)$, one determines the outward velocity and hence, the next point of collision $r_{n+1}$ and the next normal velocity $v_\perp(n+1)$. This procedure defines a Markov chain $\overline{\Phi}$ on $\overline{\Omega}=\partial \Gamma \times [0,\infty)$; denote it's transition probability kernel by $\overline{\mathcal{P}}$.

The main result of \cite{Khanin} is:
\begin{theorem} \label{thm}
The Markov Chain $\overline{\Phi}$ admits a unique invariant measure $\mu$, which is absolutely continuous with respect to Lebesgue measure. Furthermore, there exists a non-negative function $V$ on $\Omega$, as well as constants $C>0$ and $\gamma \in (0,1)$ such that
$$\sup\limits_{A \subset \overline{\Omega}}|\overline{\mathcal{P}}^n((r,v_\perp),A)-\mu(A)| \leq C \gamma^n (1+V(r,v_\perp))$$
\end{theorem}

This theorem has an important corollary:
\begin{corollary} \label{coro: convergence}
The discrete-time Markov chain $\overline{\Phi}$ is exponentially mixing and if a probability measure $\lambda$ on $\Omega$ satisfies $\int_\Omega (1+V(r,v_\perp))d\lambda < \infty$, then $\mathcal{P}_*^n \lambda$ converges to $\mu$ exponentially fast with the following bound on the rate of convergence:
$$\|\mathcal{P}^n_* \lambda - \mu \| \leq \tilde{C} \tilde{\gamma}^n \int_\Omega (1+V(r,v_\perp))d\lambda.$$
\end{corollary}

The function $V$ above is called a potential or a Lyapunov function and provides the rates with which the dynamics converges to the \textquoteleft center', represented by a level set of $V$. In \cite{Khanin} $V$ is given explicitly as follows
\begin{equation} \label{eqn:potential v_perp}
V(v_\perp) =\left\{
               \begin{array}{ll}
                 e^{\epsilon v_\perp^2}, & v_\perp>v^{\max}_\perp; \\
                 v^{-\gamma}_\perp, & v_\perp<v^{\min}_\perp;\\
                 A, & v^{\min}_\perp \leq v_\perp \leq v^{\max}_\perp.
               \end{array}
             \right.
\end{equation}
for some $0< v_\perp^{\min} < v_\perp^{\max} < \infty$, $\epsilon < \beta_{\min} = \min\{\beta_1, \cdots, \beta_p\}$ and $0<\gamma<2$. Though $\gamma$ is fixed to $1$ in \cite{Khanin}, the argument can easily be extended for any $0<\gamma<2$. Note that $V$ depends on $v_\perp$ only. A useful observation for our analysis is that, by \cite[Theorem 3.2]{Hairer}, $\int V(v_\perp)d\mu < \infty$, from which we obtain upper bounds on the density of $\mu$ near $0$.

Another important property of the discrete-time system, and the second condition for the application of Harris' ergodic theorem, is the minorization condition on compact sets. It guarantees the coupling required for fast mixing. For any $0<v_\perp^{\min} < v_\perp^{\max}<\infty$, let $C:=\{(r,v_\perp): v_\perp^{\min} \leq v_\perp \leq v_\perp^{\max}\}$ and let $m_C$ be the uniform probability measure on $C$. It was shown in \cite{Khanin} that there exist $N$ and $\eta$ such that
\begin{equation} \label{eqn: minorization}
\inf\limits_{(r,v_\perp) \in C}\mathcal{P}^{N}((r,v_\perp),\cdot) \geq \eta m_C.
\end{equation}
The minorization condition for discrete-time dynamics is used in this paper to provide lower bounds on the measures of certain sets.

A discrete-time Markov chain $\overline{\Phi}$ is called irreducible if for all $\overline{\mathbf{z}} \in \overline{\Omega}$, whenever Leb$(A)>0$, there exists some $n>0$, possibly dependent on both $\overline{\mathbf{z}}$ and $A$, such that $\mathcal{P}^n(\overline{\mathbf{z}},A)>0$; the definition for continuous-time Markov processes is similar, with $n$ replaced by $t$.

\begin{proposition} \label{prop: path}
For any two initial states $(x,v)$ and $(x',v')$ in $\Omega$, there exists a continuous time sample path from $(x,v)$ to $(x',v')$
\end{proposition}

\noindent
\textbf{Proof of Prop. \ref{prop: path}}\\
Let $(r,v_\perp)$ and $(r',v_\perp')$ be the thermostat positions and normal velocities at the collision after $(x,v)$ and at the collision before $(x',v')$, respectively. Then Prop.~3 in \cite{Khanin} guarantees that there exists a sample path between $(r,v_\perp)$ and $(r',v_\perp')$ provided that $v_\perp, v_\perp' \in [v_\perp^{\min},v_\perp^{\max}]$. However, the only use of the restriction $v_\perp, v_\perp' \in [v_\perp^{\min},v_\perp^{\max}]$ in the proof of Prop. 3 in \cite{Khanin} is for the bound on the number of collisions along such a path. Since the number of collisions is irrelevant for showing the existence of a connecting sample path, we conclude that such a path exists for all $v_\perp$ and $v_\perp'$ (see Lemmas 5 and 6 in \cite{Khanin} for details) and this trivially extends to a path between $(x,v)$ and $(x',v')$ in continuous time. $\square$

\begin{proposition} \label{prop: irreducible}
The Markov chain $\Phi_t$ is irreducible.
\end{proposition}

\noindent
\textbf{Proof of Prop. \ref{prop: irreducible}} \\
Given $\mathbf{z}=(x,v) \in \Omega$ and a set $A$ with Leb$(A)>0$, let $(x',v')$ be a Lebesgue density point of $A$. Let $(r,v_\perp)$ and $(r',v_\perp')$ be the thermostat positions and normal velocities at the collision after $(x,v)$ and at the collision before $(x',v')$, respectively, and let $U$ be an open ball around $(r',v_\perp')$. The Props. $4$ and $5$ in \cite{Khanin} imply that if $v_\perp, v_\perp' \in [v_\perp^{\min},v_\perp^{\max}]$, then there exists $n$ such that $\mathcal{P}^{n}((r,v_\perp),U)>0$. We are only interested in the positivity of $\mathcal{P}^{n}((r,v_\perp),U)$ and do not require any lower bounds, so we can drop the assumptions on $v_\perp, v_\perp'$. Since the paths from $(x,v)$ to $(r,v_\perp)$ and from $(r',v_\perp')$ to $(x',v')$ are deterministic, we conclude that there exists $t>0$ such that
$\mathcal{P}^{t}((x,v),A) >0$. $\square$

\subsection{Existence} \label{subsect: finiteness}

The result of \cite{Khanin} guarantees existence of a unique invariant probability measure for the discrete Markov chain $\overline{\Phi}$, which is absolutely continuous. Since dynamics of $\Phi_t$ is deterministic between collisions, an invariant measure for $\Phi_t$ may be constructed using a suspension flow over the discrete-time chain. This alone does not guarantee that this invariant measure would be unique or even finite. We show finiteness by estimating the integral of the roof function over the phase space. We obtain an upper bound using the potential $V$ in \cite{Khanin}. Uniqueness follows from the irreducibility of $\Phi_t$, Prop. \ref{prop: irreducible}. The argument for the uniqueness is the same as in \cite{Balint,GLP,Yarmola} and can be found in the outlines of the proofs of the main theorems in either of the papers.

For some of our future analysis it is more convenient to use a geometric description similar to that used in billiards. Namely, instead of the random variable $v_t(n)$, we shall consider a random variable $\varphi(n)$, which represents an angle relative to the normal vector at the point from which a particle leaves after collision. It is easy to show that $\varphi(n)$ is drawn from the distribution $\rho_{v_\perp(n)}(\varphi_n) d \varphi_n=\sqrt{\frac{\beta_i}{\pi}} \frac{v_\perp(n)}{\cos^2(\varphi_n)} e^{-\beta_i v^2_\perp(n) \tan^2(\varphi_n)} d \varphi_n$, which depends on the $\beta_i$ associated with $\partial D_i \ni r$ and $v_\perp(n)$. We shall also use notation $\varphi'$ for the incoming angle at the point of collision $r_{n+1}$. Obviously, $r_{n+1}$, $v_\perp(n+1)$, and $\varphi_{n}'$ are completely determined by $r_n$, $v_\perp(n)$, and $\varphi_n$. Note that $v_\perp(n+1)=\frac{\cos(\varphi_n')}{\cos(\varphi_n)}v_\perp(n)$.

Let $f(r,v_\perp)$ be the density of the unique absolutely continuous invariant measure for the discrete Markov chain $\overline{\Phi}$, i.e. $d\mu = f(r,v_\perp)dr dv_\perp$. Before constructing the suspension flow, we need to re-introduce a third variable $\varphi$. The invariant measure for the discrete process with $(r, v_\perp, \varphi)$ variables is
$$d \mu_1=f(r,v_\perp)\rho_{v_\perp}(\varphi)dr dv_\perp d \varphi= f(r,v_\perp) \sqrt{\frac{\beta_i}{\pi}} \frac{v_\perp}{\cos^2(\varphi)} e^{- \beta_i v^2_\perp \tan^2(\varphi)}dr dv_\perp d \varphi.$$

Let $\nu$ by the invariant measure for $\Phi_t$ obtained by constructing a suspension flow over the discrete dynamics in a standard way. See \cite[p.~31]{Chernov} for details and additional references. The suspension flow is defined on a $4$-dimensional space in variables $(r,v_\perp,\varphi,t)$, which we would still refer to as $\Omega$.

Let $\sigma(r,\varphi)$ be the distance of flight between two thermostats that originates at $(r,\varphi)$. If, in addition, the flight originates with a given $v_\perp$, then the speed during the flight is $\frac{v_\perp}{\cos(\varphi)}$.
In order to show that this invariant measure is finite, we need to show that the time of flight $\frac{\sigma(r,\varphi)\cos(\varphi)}{v_\perp}$ is integrable with respect to $d\mu_1$,

\begin{equa}
\int \frac{\sigma(r,\varphi)\cos(\varphi)}{v_\perp} d\mu_1 & = \int \frac{\sigma(r,\varphi)\cos(\varphi)}{v_\perp} f(r,v_\perp) \sqrt{\frac{\beta_i}{\pi}} \frac{v_\perp}{\cos^2(\varphi)} e^{- \beta_i v^2_\perp \tan^2(\varphi)}dr dv_\perp d \varphi \\
& \leq \sigma_{\max} \int f(r,v_\perp) \sqrt{\frac{\beta_i}{\pi}} \frac{1}{\cos(\varphi)} e^{- \beta_i v^2_\perp \tan^2(\varphi)} d \varphi dr dv_\perp \\
& \leq \sigma_{\max} \sqrt{\frac{\beta_{\max}}{\pi}} \int f(r,v_\perp) \cos(\varphi(w)) e^{- \beta_i v^2_\perp w^2} d w dr dv_\perp \\
& \leq \sigma_{\max} \sqrt{\frac{\beta_{\max}}{\pi}} \int \frac{f(r,v_\perp)}{v_\perp} drdv_\perp = \sigma_{\max} \sqrt{\frac{\beta_{\max}}{\pi}} \int \frac{1}{v_\perp} d\mu.
\end{equa}

Here $\sigma_{\max}$ is the maximum distance of flight and $\beta_{\max}=\max\{\beta_1, \cdots, \beta_p\}$.

The potential $V$ introduced in \cite{Khanin} was $V(v_\perp)=v_\perp^{-1}$ for $v_\perp < v_\perp^{\min}$ for some small $v_\perp^{\min}$. By Theorem 3.2 in \cite{Hairer} we know that $\int V(r,v_\perp)d\mu < \infty$.
We conclude that $\int \frac{1}{v_\perp} d\mu \leq \int V(r,v_\perp)d\mu < \infty$. Therefore,

$$\int \frac{\sigma(r,\varphi)\cos(\varphi)}{v_\perp} d\mu_1 < \infty$$

and the invariant measure for the suspension flow is indeed finite. $\square$

\subsection{Mixing} \label{subsect: ergodicity}

The proof of mixing and convergence of initial distributions to the invariant measure for $\Phi_t$ relies on sampling the Markov process $\Phi_t$ at integer times, which generates a discrete-time skeleton chain $\Phi_1$.  Denote the transition probability kernel for $\Phi_1$ by $\mathcal{P}_1$. The following theorem by Meyn and Tweedie relates skeleton chains to the ergodicity of the Markov processes.

\begin{definition}
The Markov process $\Phi_t$ is called ergodic if an invariant probability measure $\nu$ exists and
$$\lim\limits_{t \to \infty} \|\mathcal{P}^t(\mathbf{z},\cdot)-\nu\|=0, \;\;\;\;\; \forall \mathbf{z} \in \Omega$$
\end{definition}

\begin{theorem} \cite{MeynII} \label{thm: Meyn}
Suppose the Markov process $\Phi_t$ is irreducible and $\nu$ is an invariant probability measure for $\Phi_t$. Then $\Phi_t$ is ergodic if and only if there exists a skeleton chain which is irreducible.
\end{theorem}

The mixing of the invariant measure for the Markov process $\Phi_t$ and the convergence of initial distributions to the invariant measure follow from ergodicity by the Dominated Convergence Theorem.
Indeed,
\begin{equa}
\lim\limits_{t \to \infty}\|\mathcal{P}^t\lambda - \nu\| & =\lim\limits_{t \to \infty} \sup\limits_{A \subset \Omega}|\int_\Omega (\mathcal{P}^t(
\mathbf{z},A)-\nu(A))d\lambda| \\
& \leq \lim\limits_{t \to \infty} \int_\Omega \|P^t(\mathbf{z},\cdot)-\nu(\cdot)\|d\lambda= 0,
\end{equa}
and
\begin{equa}
\lim\limits_{t \to \infty} \sup\limits_{B \in \mathcal{B}} |\int_A \mathcal{P}^t(\mathbf{z},B)d\nu - \nu(A)\nu(B)| & \leq  \lim\limits_{t \to \infty} \sup\limits_{B \in \mathcal{B}} \frac{1}{\nu(A)}|\int_A \mathcal{P}^t(\mathbf{z},B)d\nu -\nu(A)\nu(B)| \\
& = \lim\limits_{t \to \infty} \sup\limits_{B \in \mathcal{B}} |\int_\Omega \mathcal{P}^t(\mathbf{z},B)\frac{\mathbf{1}_A}{\nu(A)}d\nu -\nu(B)| \\
& = \lim\limits_{t \to \infty}\|\mathcal{P}^t\lambda - \nu\| = 0,
\end{equa}
where $d\lambda = \frac{\mathbf{1}_A}{\nu(A)}d\nu$.

Ergodicity follows once we show that the time-$1$ sampled chain $\Phi_1$ is irreducible. Prop. \ref{prop: path} guarantees that for any two initial states $(x,v)$ and $(x',v')$ in $\Omega$, there exists a continuous time sample path from $(x,v)$ to $(x',v')$.

We must now show that such a sample path from Prop. \ref{prop: path} can be modified to last integer time. Without loss of generality we can assume that each of the continuous-time paths from Prop. \ref{prop: path} passes through a given $(x_0,v_0)$. If there exists a family of closed sample paths from $(x_0,v_0)$ to $(x_0,v_0)$ such that the fractional parts of the times they use cover the interval $[0,1]$, then a particle can follow these sample paths to make the total time of traversing the path integer. By continuity, there exists a range of paths of this kind covering the time interval $[t_1,t_2]$ for some $t_1<t_2$. Then, if we repeat each of these paths $k$ times, the time interval of the $k^{th}$ iteration covers $[kt_1,kt_2]$. If $k$ is large enough, $kt_2-kt_1>1$, which means that all possible fractional parts of the times to complete such iterated paths cover $[0,1]$. By adding an appropriate path of this kind to the existing path from $(x,v)$ to $(x',v')$ through $(x_0,v_0)$, we produce a sample path taking integer amount of time to complete.

Similar to proof of Prop.~\ref{prop: irreducible}, by modifying Props.~5 and 6 in \cite{Khanin} for any $v_\perp,v_\perp'$ while dropping lower bounds on the densities, we conclude that for all $(x,v)$ and $(x',v')$, there exists $n$ and an open ball $U_{(x',v')} \ni (x',v')$ such that $\mathcal{P}_1^n((x,v),\cdot) $ has strictly positive density on $U_{(x',v')}$. It follows that the time-$1$ chain $\Phi_1$ is irreducible. $\square$

\subsection{Mixing is not exponential} \label{subsect: not exponential}

The main reason for sub-exponential rates of mixing is that, occasionally, the particle acquires low speed due to the randomness of the thermostats, which leads to long traveling times between collisions. Let $B_\tau$ be the set of all $\mathbf{z} \in \Omega$ such that the particle will not collide with $\partial \Gamma$ before time $\tau$. Denote the invariant measure by $\nu$. We show that the $\nu$-measure of the set $B_\tau$ is bounded above and below by $\frac{\xi}{\tau^2} \leq \nu(B_\tau) \leq \frac{\xi'}{\tau^{\gamma}}$ for any $0 < \gamma < 2$ and some $\xi, \xi'(\gamma)>0$. The estimate is derived using the information about the system gathered in the proof of the main theorem in $\cite{Khanin}$. Then the dynamics of our system is statistically very similar to the dynamics of an expanding map with a neutral fixed point (a.k.a. Pomeau-Manneville).  Indeed, the mass originally in $B_\tau$ evolves deterministically for time $\tau$; extra mass is \textquoteleft deposited' from the parts of the phase space where particles experience nearly tangential collisions (which reduces the speed). Using arguments similar to \cite{LSY}, we obtain:

\begin{proposition} \label{prop: nonexp mixing}
For any probability distribution $\lambda \ll \nu$ such that $d\lambda \geq (1+c)d\nu$ on $B_{\tau_0}$ for some $c>0$ and $\tau_0>0$ large enough, $\mathcal{P}^\tau_* \lambda$ converges to $\nu$ with sub-exponential rate. More precisely, there exists $\varsigma>0$ such that for an increasing sequence $\{\tau_j\}$,
$$\|\mathcal{P}^{\tau_j}_* \lambda - \nu\| \geq \frac{\varsigma}{\tau_j^2}.$$
In particular, the unique invariant measure $\nu$ for the Markov Process $\Phi_t$ is not exponentially mixing, i.e., there exists a Borel set $A \in \Omega$ such that
$$\sup\limits_{B \in \mathcal{B}}|\int\limits_{A}\mathcal{P}^\tau_j(\mathbf{z},B)d\nu-\nu(A)\nu(B)| \geq  \nu(A) \cdot \frac{\varsigma}{2\tau_j^2}.$$
\end{proposition}

\medbreak

\textbf{Proof of Prop.~\ref{prop: nonexp mixing}.}
Let us first consider the equilibrium situation, i.e., when $\beta=\beta_1= \cdots=\beta_p$. Then

$$d \nu =c \sqrt{\frac{\beta}{\pi}} \frac{v^2_\perp}{\cos^2(\varphi)} e^{-\beta v^2_\perp /\cos^2(\varphi)} d v_\perp dr d \varphi dt$$
$$= c s^2 e^{-\beta s^2} \cos(\varphi) ds dr d \varphi dt,$$
\noindent
where $s=\frac{v_\perp}{\cos(\varphi)}$ is the speed of the particle. We would like to estimate $\nu(B_\tau)$. The particle with speed $s$ will not reach a thermostat in time $\tau$ if $t \in [0,\frac{\sigma(r,\varphi)}{s}-\tau)$, where $\sigma(r,\varphi)$ denotes the flight distance to the next thermostat is the particle originates at $(r,\varphi)$. Thus
\begin{equa}
\nu(B_\tau) & =  \int\limits_{\partial \Gamma} \int\limits_{-\frac{\pi}{2}}^{\frac{\pi}{2}} \int\limits_{0}^{\frac{\sigma(r,\varphi)}{\tau}}  \int\limits_{0}^{\frac{\sigma(r,\varphi)}{s} - \tau} c s^2 e^{-\beta s^2} \cos(\varphi) dt ds d \varphi dr \\
& =\int\limits_{\partial \Gamma} \int\limits_{-\frac{\pi}{2}}^{\frac{\pi}{2}} \int\limits_{0}^{\frac{\sigma(r,\varphi)}{\tau}} c (\frac{\sigma(r,\varphi)}{s} - \tau) s^2 e^{-\beta s^2} \cos(\varphi) ds d \varphi dr \\
& =\int\limits_{\partial \Gamma} \int\limits_{-\frac{\pi}{2}}^{\frac{\pi}{2}} c[\frac{\sigma(r,\varphi)}{2 \beta}-\frac{\sqrt{\pi}\tau \mbox{Erf}[\frac{\sqrt{\beta}\sigma(r,\varphi)}{\tau}]}{4 \beta^{\frac{3}{2}}}]\cos(\varphi)d\varphi dr \\
& =\int\limits_{\partial \Gamma} \int\limits_{-\frac{\pi}{2}}^{\frac{\pi}{2}} c[\frac{\sigma(r,\varphi)^3}{6 \tau^2}+O(
\frac{1}{\tau^4})]\cos(\varphi)d\varphi dr \approx \frac{1}{\tau^2},
\end{equa}
\noindent
where $f \approx g$ means $f(\tau)=\Theta(g(\tau))$, i.e., there exists $\zeta, \zeta'>0$ such that $\zeta g(\tau) \leq f(\tau) \leq \zeta' g(\tau)$, for all $\tau$.

We show that a similar estimate holds in the non-equilibrium setting.

\begin{lemma} \label{lemma: slow mass}
There exist $\delta>0$ and $\tau_0$ large enough, such that for any $\gamma$, $0<\gamma<2$, there exist $\xi,\xi'>0$ such that for $k\delta>\tau_0$, $\frac{\xi}{(k\delta)^2} \leq \nu(B_{k\delta}) \leq \frac{\xi'}{(k\delta)^{\gamma}}$.
\end{lemma}

Assuming Lemma \ref{lemma: slow mass}, we complete the proof of Prop. \ref{prop: nonexp mixing} using an argument similar to \cite{LSY}.

Let $\lambda \ll \nu$ with $d\lambda = \varphi d\nu$ be such that $\varphi \geq 1+c$ on $B_{\tau_0}$ for some $c>0$ and $\tau_0>0$. Then for $k\delta>\tau_0$
\begin{equa}
\|\mathcal{P}^{n\delta}_* \lambda - \nu\| & \geq (\mathcal{P}^{n\delta}_* \lambda)(B_{k\delta}) - \nu(B_{k\delta}) \geq \lambda(B_{k\delta+n\delta})-\nu(B_{k\delta}) \\
       & \geq (1+c)\nu(B_{k\delta+n\delta})-\nu(B_{k\delta}) = [(1+c)\frac{\nu(B_{k\delta+n\delta})}{\nu(B_{k\delta})}-1]\nu(B_{k\delta}) \\
\end{equa}
Let $k=Nn$ for some $N$. If, for an infinite sequence $n_j$, 
$$[(1+c)\frac{\nu(B_{(N+1)n_j\delta})}{\nu(B_{Nn_j\delta})}-1] \geq \frac{c}{2},$$ 
then
\begin{equa} \label{eqn: measure ratio}
\|\mathcal{P}^{n_j\delta}_* \lambda - \nu\| \geq \frac{c}{2}\frac{\xi}{N^2 n_j^2 \delta^2}=:\frac{\varsigma}{\tau_j^2} \\
\end{equa}
where $\tau_j=n_j\delta$. Thus the convergence of $\mathcal{P}^\tau_* \lambda$ to $\nu$ is sub-exponential.

It remains to show that there exists $N$ such that \ref{eqn: measure ratio}, which is equivalent to $\frac{\nu(B_{(N+1)n\delta})}{\nu(B_{Nn\delta})} \geq \frac{2+c}{2(1+c)}  =: \lambda$, holds for infinitely many $n$. 

For each $N$, let $A_N=\{n: \nu(B_{(N+1)n\delta}) \geq \lambda \nu(B_{Nn\delta})\}$. Assume $A_N$ is finite for all $N$. Let $m_N=\sup\{A_N\}+1$ if $A_N \ne \emptyset$ and $m_N=1$ otherwise. Then for any $N$ and $\forall n \geq m_N$, $\nu(B_{(N+1)n\delta}) < \lambda \nu(B_{Nn\delta})$. For given $N$, let us see how much drop one can accumulate by applying this property on non-intersecting intervals $[Nn_l\delta,(N+1)n_l\delta)$ for appropriately chosen $n_l$.

Let $n_0=m_N$. Let $n_1$ be such that $Nn_1 > (N+1)m_N$, e.g., let $n_1 = \lceil \frac{N+1}{N}m_N \rceil \leq \frac{N+1}{N}m_N +1$. Let $n_l = \lceil \frac{N+1}{N}n_{l-1} \rceil \leq \frac{N+1}{N}n_{l-1}+1$. Then
\begin{equa}
n_l & \leq \left(\frac{N+1}{N}\right)^l m_N + \left(\frac{N+1}{N}\right)^{l-1}+ \cdots + \frac{N+1}{N} +1 \\
& = \left(\frac{N+1}{N}\right)^l m_N + \left(\frac{N+1}{N}\right)^l N -N \leq \left(\frac{N+1}{N}\right)^l (m_N + N),
\end{equa}
and
$$\lambda^l > \frac{\nu(B_{Nn_l\delta})}{\nu(B_{Nm_N\delta})} \geq \frac{\xi N^{\gamma} m_N^{\gamma} \delta^{\gamma}}{\xi' N^2 (\frac{N+1}{N})^{2l} (m_N+N)^2 \delta^2}=\frac{\xi m_N^{\gamma}}{\xi' \delta^{2-\gamma} N^{2-\gamma} (m_N+N)^2}\left(\frac{N}{N+1}\right)^{2l}.$$

Choose $N$ such that $\frac{N}{N+1}>\sqrt{\lambda}$. Then choose $l$ such that $(\frac{\xi m_N^{\gamma}}{\xi' \delta^{2-\gamma} N^{2-\gamma} (m_N+N)^2})^{1/l}>\sqrt{\lambda}$. Then we obtain $\lambda^l > \lambda^{l/2} \lambda^{l/2} = \lambda^l$, a contradiction. Therefore, there exists $N$ such that $A_N := \{n_i\}$ is infinite.

\medbreak
To obtain a lower bound on the rate of mixing, it suffices to pick $A=B_{\tau_0}$ and $\lambda \ll \nu$ such that $d\lambda = (\mathbf{1}_A/\nu(A))d\nu$. Then
$$\sup\limits_{B \in \mathcal{B}}\frac{1}{\nu(A)}|\int\limits_{A}\mathcal{P}^\tau_j(\mathbf{z},B)d\nu-\nu(A)\nu(B)| = \|\mathcal{P}^{n_j\delta}_* \lambda - \nu\| \geq \frac{\varsigma}{2\tau_j^2}.$$
\begin{flushright}
$\square$
\end{flushright}

\medbreak

\textbf{Proof of Lemma \ref{lemma: slow mass}.}
Let us first obtain an upper bound on $\nu(B_\tau)$. Denote by $f(r,v_\perp)$ the density of the invariant measure $\mu$ for the discrete dynamics, i.e., $d\mu=f(r,v_\perp)dr dv_\perp$.  Denote the maximal particle's flight distance by $\sigma_{\max}$. Then the maximal possible speed in $B_\tau$ is $\frac{\sigma_{\max}}{\tau} \geq \frac{v_\perp}{\cos(\varphi)}$, implying that $v_\perp \leq \frac{\sigma_{\max}}{\tau}$. Thus

\begin{equa}
\nu(B_\tau) & \leq  \int\limits_{\partial \Gamma} \int\limits_{0}^{\frac{\sigma_{\max}}{\tau}} \int\limits_{-\frac{\pi}{2}}^{\frac{\pi}{2}} \int\limits_{0}^{\sigma_{\max}} c f(r,v_\perp) \rho_{v_\perp}(\varphi) dt d\varphi d v_\perp dr\\
& \leq \sigma_{\max} \int\limits_{\partial \Gamma} \int\limits_{0}^{\frac{\sigma_{\max}}{\tau}} \int\limits_{-\frac{\pi}{2}}^{\frac{\pi}{2}} c f(r,v_\perp) \sqrt{\frac{\beta_i}{\pi}} \frac{v_\perp}{\cos^2(\varphi)} e^{-\beta_i v_\perp^2 \tan^2(\varphi)} d\varphi d v_\perp dr \\
& \leq K \int\limits_{\partial \Gamma} \int\limits_{0}^{\frac{\sigma_{\max}}{\tau}} \int\limits_{-\infty}^{\infty} v_\perp f(r,v_\perp) e^{-\beta_i v_\perp^2 w^2} dw d v_\perp dr = K \int\limits_{\partial \Gamma} \int\limits_{0}^{\frac{\sigma_{\max}}{\tau}} f(r,v_\perp) dv_\perp dr \\
& =  K \int\limits_{\partial \Gamma} \int\limits_{0}^{\frac{\sigma_{\max}}{\tau}} v_\perp^{\gamma} \frac{f(r,v_\perp)}{v_\perp^\gamma} dv_\perp dr \leq K \left( \frac{\sigma_{\max}}{\tau}\right)^\gamma \int\limits_{\partial \Gamma} \int\limits_{0}^{\frac{\sigma_{\max}}{\tau}} \frac{f(r,v_\perp)}{v_\perp^\gamma} dv_\perp dr \\
& \leq K \left(\frac{\sigma_{\max}}{\tau}\right)^\gamma \int V(r,v_\perp)d\mu.
\end{equa}
\noindent
since $V(r,v_\perp)=1/v_{\perp}^\gamma$ for $0<\gamma<2$ and $v_\perp \leq v_\perp^{\min}$. The last integral is finite by Theorem 3.2 in \cite{Hairer}. This establishes an upper bound for Lemma \ref{lemma: slow mass}.

\medbreak

We will obtain a lower bound using the following lemma:

\begin{lemma} \label{lemma: v_perp lower bound}
Let $\mu_1$ be the lift of $\mu$ to the $(r,v_\perp,\varphi)$-space and let $A(\overline{v}_\perp):=\{(r,v_\perp,\varphi): v_\perp^{\min} \leq v_\perp \leq v_\perp^{\max}, \;  v_\perp' \leq \overline{v}_\perp \}$, where
%and $0<v_\perp^{\max}<v_\perp^{\max}<\infty$  are some constants such that the potential and minorization conditions in \cite{Khanin} are satisfied and
$v_\perp'$ denotes the normal velocity of the particle upon its next collision given that it originates at $(r,v_\perp,\varphi)$. Then there exists $\overline{v}_\perp^0>0$ such that
$$\mu_1(A(\overline{v}_\perp)) \geq \kappa \overline{v}_\perp^3$$ for $\overline{v}_\perp<\overline{v}_\perp^0$ and some $\kappa>0$.
\end{lemma}

We will prove Lemma \ref{lemma: v_perp lower bound} at the end of this subsection.

Note that if $\overline{v}_\perp^0 \ll v_\perp^{\min}$ and  $(r,v_\perp,\varphi) \in A(\overline{v}_\perp)$, then $\varphi$ is uniformly bounded above by some $\vartheta_0>0$ and the time of flight between thermostats is bounded below by $\sigma_{\min} \cos(\vartheta_0)/v_\perp^{\max}$. Fix $\delta<\sigma_{\min} \cos(\vartheta_0)/v_\perp^{\max}$.

Let $\tilde{A}(\overline{v}_\perp)=\{(r,v_\perp,\varphi,t) \in \Omega: (r,v_\perp,\varphi) \in A(\overline{v}_\perp)\}$. Assume at time $0$ a particle originates at $(r,v_\perp,\varphi,t) \in \tilde{A}(\overline{v}_\perp)$ and is going to collide with a thermostat within time $\delta$. By Lemma \ref{lemma: v_perp lower bound}, the $\nu$-measure of such particles is greater or equal to $\tilde{c} \delta \kappa \overline{v}_\perp^3$ for some $\tilde{c}>0$. Assume that at the next collision, the new angle $\varphi_{new}$ drawn from $\rho_{v_\perp'}(\varphi_{new})d\varphi_{new}$ satisfies $|\varphi_{new}|<\frac{\pi}{3}$. Then the particle's speed after the collision is $s' = \frac{v_\perp'}{\cos(\frac{\pi}{3})}<2 \overline{v}_\perp$. Fix $\overline{v}_\perp = \frac{\sigma_{\min}}{2(\tau+\delta)}$. Then, after the first collision, the particle will not collide with the next thermostat in time $\tau+\delta$ and, at time $\delta$, the coordinates of the particle would belong to $B_\tau$.

Thus, the $\nu$-measure of states that get to $B_\tau$ in time $\delta$ is greater or equal to
$$\mathbb{P}(|\varphi_{new}|<\frac{\pi}{3}) \times \nu(\tilde{A}(\overline{v}_\perp)) \geq \tilde{c}' \delta \kappa \overline{v}_\perp^3 = \tilde{c}' \delta \kappa \left(\frac{\sigma_{\min}}{2(\tau+\delta)}\right)^3 \geq \frac{\zeta}{\tau^3}$$

Thus $\nu(B_\tau) \geq \nu(B_{\tau+\delta})+ \frac{\zeta}{\tau^3}$ and $\nu(B_\tau) \geq \sum\limits_{k=0}^\infty \frac{\zeta}{(\tau+k\delta)^3} \approx \frac{1}{\tau^2}$. $\square$

\medbreak

\textbf{Proof of Lemma \ref{lemma: v_perp lower bound}.}

Let $C:=\{(r,v_\perp): v_\perp^{\min} \leq v_\perp \leq v_\perp^{\max}\}$, and let $m_C$ be the uniform probability measure on $C$. The minorization condition (\ref{eqn: minorization}) states that: $\exists N \in \mathbb{N}$ and $\eta>0$ such that
$$\inf\limits_{(r,v_\perp) \in C}\mathcal{P}^{N}((r,v_\perp),\cdot) \geq \eta m_C.$$
Therefore, by invariance, $\mu$ has density at least $\mu(C) \eta$ on $C$. Note that $\mu(C)>0$ since the chain is irreducible.
\medbreak

Given $(r,v_\perp)$, we would like to estimate the probability that $v_\perp'=\frac{\cos(\varphi')}{\cos(\varphi)}v_\perp \leq \overline{v}_\perp$. Using the same parametrization as in subsubsection 4.2.1 in \cite{Khanin} we find that
$\frac{\cos^2(\varphi')}{\cos^2(\varphi)}$ is a piecewise quadratic function when viewed in $w=\tan(\varphi)$-coordinates. We are interested in integrating the angle density $\rho_{v_\perp}(\varphi)$ on the intervals $(w_0, w_1):=(\tan(\varphi_0),\tan(\varphi_1))$ on which $0 \leq \frac{\cos(\varphi')}{\cos(\varphi)} \leq \frac{\overline{v}_\perp}{v_\perp}$.

$$\mathbb{P}(v_\perp' \leq \overline{v}_\perp)=\int\limits_{\frac{\cos(\varphi')}{\cos(\varphi)} \leq \frac{\overline{v}_\perp}{v_\perp} }\sqrt{\frac{\beta_i}{\pi}}\frac{v_\perp}{\cos^2(\varphi)} e^{-\beta_i v_\perp^2 \tan^2(\varphi)}d\varphi=\int\limits_{\frac{\cos(\varphi')}{\cos(\varphi)} \leq \frac{\overline{v}_\perp}{v_\perp} }\sqrt{\frac{\beta_i}{\pi}}v_\perp e^{-\beta_i v_\perp^2 w^2}dw$$
$$=\Sigma(\mbox{Erf}[\sqrt{\beta_i} v_\perp w_1]-\mbox{Erf}[\sqrt{\beta_i} v_\perp w_0])/2,$$
where $\Sigma$ denotes the sum over the piecewise components.

\begin{figure}
  \centering
  \resizebox{0.75\textwidth}{!}{
  \includegraphics{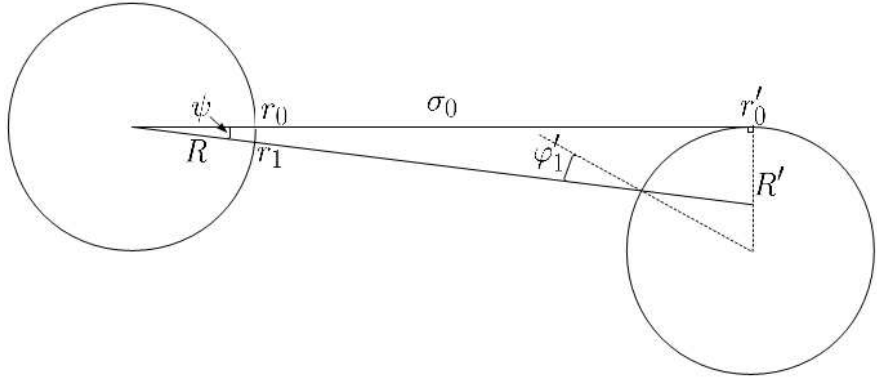}}
  \caption{Geometric locations that produce small $v_\perp'$}\label{fig:r small v_perp}
\end{figure}

The difference in the error functions is only substantial when $\sqrt{\beta_i} v_\perp w_1$ and $\sqrt{\beta_i} v_\perp w_0$ are sufficiently close to $0$. In particular, this happens near geometric locations $r \in \partial \Gamma$ from which a radially emitted particle collides with the next thermostat tangentially. See Fig. \ref{fig:r small v_perp}. We will show that, for small enough $\overline{v}_\perp$, the measure of $\{r:$ such that $w_0 \leq 0 \leq w_1$ (or $w_1 \leq 0 \leq w_0$)$\}$ is $\geq \iota (\overline{v}_\perp)^2$ for some $\iota>0$.

Let $r_0$ and $r_0'$ be such that if a particle originates radially from $r_0$ (with $\varphi_0=0$), then it arrives to $r_0'$ tangentially (i.e. $\varphi_0'=\pm \frac{\pi}{2}$). Note that $\varphi_0=0$ corresponds to $w_0=0$. Now we can start shifting $r$ keeping $w_0 \leq 0 \leq w_1$ (or $w_1 \leq 0 \leq w_0$) until we get $w_1=0$, i.e. $\cos(\varphi_1')=\frac{\overline{v}_\perp}{v_\perp^{\max}}$. The relation between $\psi=\frac{\Delta r}{R}$ (see Fig. \ref{fig:r small v_perp}) and $\varphi_1'$ is as follows:
$$\frac{R_1-\sigma_0 \tan(\psi)}{\sin(\varphi_1')}=\frac{R_1}{\cos(\psi)} \;\;\; \Leftrightarrow \;\;\; R_1 \cos(\psi)-\sigma_0 \sin(\psi)=R_1 \sin(\varphi_1')$$
Solving for $\sin(\psi)$ we get
\begin{equa}
\sin(\psi) & = \frac{-\sigma_0 R_1 \sin(\varphi_1')+R_1 \sigma_0 \sqrt{1+\frac{R_1^2}{\sigma_0^2}\cos^2(\varphi_1')}}{\sigma_0^2 + R_1^2} \\
& =\frac{-\sigma_0 R_1 \sqrt{1-(\frac{\overline{v}_\perp}{v_\perp^{\max}})^2}+R_1 \sigma_0 \sqrt{1+\frac{R_1^2}{\sigma_0^2}(\frac{\overline{v}_\perp}{v_\perp^{\max}})^2}}{\sigma_0^2 + R_1^2} \\
& \geq \frac{\sigma_0 R_1}{\sigma_0^2 + R_1^2}\frac{\min\{\frac{R_1^2}{\sigma_0^2},1\}}{4}(\frac{\overline{v}_\perp}{v_\perp^{\max}})^2 =: \iota (\overline{v}_\perp)^2
\end{equa}

From computations in subsubsection 4.2.1 in \cite{Khanin} one also obtains that $\frac{R \overline{v}_\perp}{2 \sigma_0 v_\perp} \leq |w_1-w_0| \leq  \frac{R \overline{v}_\perp}{ \sigma_0 v_\perp}$. Therefore, for $\overline{v}_\perp$ is small enough, if $r$ is chosen such that $w_0 \leq 0 \leq w_1$ (or $w_1 \leq 0 \leq w_0$) and $v_\perp^{\min} \leq v_\perp \leq v_\perp^{\max}$,
$$\mathbb{P}(v_\perp' \leq \overline{v}_\perp) \geq \mathrm{Erf}[\sqrt{\beta_i} v_\perp w_1]-\mathrm{Erf}[\sqrt{\beta_i} v_\perp w_0])/2 \geq \mathrm{Erf}[\sqrt{\beta_i} v_\perp |w_1-w_0|]/2 \geq \sqrt{\frac{\beta}{\pi}}\frac{R \overline{v}_\perp}{4 \sigma_0}$$

Therefore, $\mu_1(A(\overline{v}_\perp)) \geq \mu(C) \eta \iota (\overline{v}_\perp)^2 \eta \sqrt{\frac{\beta}{\pi}}\frac{R \overline{v}_\perp}{4 \sigma_0}=:\kappa \overline{v}_\perp^3$. $\square$

\medbreak

\ack The author would like to thank the unnamed referees for valuable comments and suggestions as well as for pointing out relevant papers unknown to the author; Lai-Sang Young for enlightening discussions; Jean-Pierre Eckmann and Noe Cuneo for valuable comments, constructive criticism, and reading the drafts; and Michael Milligan for proof-reading, encouragement, nurture, inspiration and support. The author was partially supported by the ERC Advanced Grant "Bridges"
\medbreak

\end{document}